\documentclass[letterpaper,twocolumn,english,aps,prl,floatfix, showpacs, amsfonts, amssymb]{revtex4}
\usepackage[T1]{fontenc}
\usepackage[latin1]{inputenc}
\usepackage{amsmath}
\usepackage{graphicx}
\usepackage{amssymb}
\usepackage{times}

\makeatletter
\usepackage{amscd}
\usepackage{bm}

\usepackage{babel}
\makeatother
\begin{document}

\title{Random Antiferromagnetic $SU(N)$ Spin Chains }

\author{José Abel Hoyos}

\email{joseabel@ifi.unicamp.br}

\affiliation{Instituto de F\'{\i}sica Gleb Wataghin, Unicamp, Caixa Postal 6165,
13083-970 Campinas, São Paulo, Brazil}

\author{E. Miranda}

\email{emiranda@ifi.unicamp.br}

\affiliation{Instituto de F\'{\i}sica Gleb Wataghin, Unicamp, Caixa Postal 6165,
13083-970 Campinas, São Paulo, Brazil}

\date{\today{}}

\begin{abstract}
We analyze random isotropic antiferromagnetic $SU\left(N\right)$
spin chains using the real space renormalization group. We find that
they are governed at low energies by a universal infinite randomness
fixed point \emph{different} from the one of random spin-$\frac{1}{2}$
chains. We determine analytically the important exponents: the energy-length
scale relation is $\Omega\sim\exp\left(-L^{\psi}\right)$, where $\psi=1/N$,
and the mean correlation function is given by $\overline{C_{ij}}\sim\left(-1\right)^{i-j}/\left|i-j\right|^{\phi}$,
where $\phi=4/N$. Our analysis shows that the infinite-$N$ limit
is unable to capture the behavior obtained at any finite $N$.
\end{abstract}


\maketitle
The identification of several universality classes of disordered quantum
systems in one dimension (1D) has seen enormous progress in recent
years. Prominent among those is the case of random antiferromagnetic
spin-$\frac{1}{2}$ chains, which have been shown to be controlled
by an infinite randomness fixed point (IRFP) at low energies \cite{madasguptahu,fisherrandomchain}.
Many properties of this so-called random singlet phase have been obtained,
e. g., the spin susceptibility $\chi\sim1/T\log^{2}T$, and the spin
correlation function $C_{ij}=\left\langle \mathbf{S}_{i}\cdot\mathbf{S}_{j}\right\rangle $
is such that its mean value $\overline{C_{ij}}\sim\left(-1\right)^{i-j}/\left|i-j\right|^{2}$,
while the typical one $\left|C_{ij}\right|_{typ}\sim\exp(-\sqrt{\left|i-j\right|})$.
Further studies of 1D spin systems have uncovered a wide variety of
behaviors such as various Griffiths phases \cite{fishertransising,hymanetal,yangetal,melinetal,yusufyang1,saguiaetal,hoyosmiranda1}
and large spin phases \cite{westerbergetal,yusufyang2,hoyosmiranda1}.

It is the purpose of this paper to extend these analyses by enlarging
the symmetry group from $SU\left(2\right)$ to $SU\left(N\right)$.
We have several motivations for this. The inclusion of orbital degrees
of freedom often leads to the enlargement $SU\left(2\right)\to SU\left(N\right)$.
The strong spin-orbit interaction in rare-earth elements locks spin
and orbital moments into a large multiplet with degeneracy $N$, whose
description requires the enlargement from $SU\left(2\right)$ to $SU\left(N\right)$
\cite{coqblinschrieffer}. Recently, a realization of a self-conjugate
$SU(4)$ spin chain has been proposed in a pillar array of semiconducting
quantum dots, where the symmetry-breaking effect of the intra-dot
electron-electron interaction is minimized due to the peculiarities
of the dot potential \cite{onufrievmarston}. Several other possible
realizations of enlarged symmetry have been considered in the literature
(see, for example, Ref.~\cite{azariaetal}). In any of these cases,
the effects of disorder are clearly of interest. Furthermore, the
large-$N$ limit of $SU(N)$ spin models is of considerable interest.
In this limit, many models can be solved by saddle-point methods and
$1/N$ corrections can be obtained in a controlled manner \cite{sachdevbook}.
The hope behind this approach is that the physics of $N=2$ is at
least qualitatively captured as $N\to\infty$. The solution of a sequence
of models \emph{as a function of} $N$, though rarely possible, can
determine the validity of this idea. We will show that the random
antiferromagnetic $SU\left(N\right)$ chain provides just such a solution.
Interestingly, the low-energy physics at finite $N$ is never captured
at infinite $N$. We will show, however, that for any finite $N$,
the system is governed by a \emph{new universal IRFP} with characteristic
exponents which we calculate and depend only on the group rank.

We will focus on the following Hamiltonian\begin{equation}
H=\sum_{i=1}^{L}J_{i}\bm{\Gamma}_{i}\cdot\bm{\Gamma}_{i+1},\label{eq:ham}\end{equation}
 where $J_{i}$ are positive independent random variables distributed
according to $P_{0}(J)$, and the components of $\bm{\Gamma}_i$,
$\Gamma_{i}^{a}$ ($a=1,\ldots,N^{2}-1$), 
are the generators of a representation of $SU(N)$. We will confine
our analysis to totally antisymmetric irreducible representations.
These correspond to Young tableaux with one column and $Q_{i}$ lines
\cite{jones}. They are conveniently expressed with the help of auxiliary
fermionic operators $c_{i\alpha}^{\phantom{\dagger}}$ ($\alpha=1,\ldots,N$)
through $\Gamma_{i}^{a}=c_{i\alpha}^{\dagger}\Gamma_{\alpha\beta}^{a}c_{i\beta}^{\phantom{\dagger}}$,
where $\Gamma_{\alpha\beta}^{a}$ are the generator matrices of the fundamental
representation of $SU(N)$. The fermions obey the constraint $\sum_{\alpha=1}^{N}c_{i\alpha}^{\dagger}c_{i\alpha}^{\phantom{\dagger}}=Q_{i}$.
We considered the cases where the $Q_{i}$'s are random and $Q_{i}=Q=\mathrm{const.}$

To treat the Hamiltonian (\ref{eq:ham}), we generalize the real-space
renormalization group method introduced by Ma, Dasgupta and Hu \cite{madasguptahu}.
Our generalization is reminiscent of the treatment of random ferro-
and antiferromagnetic spin chains \cite{westerbergetal}. We first
find the largest bond energy of the system, $\Omega=\max\left\{ \Delta_{i}\right\} $.
We define $\Delta_{i}$ as the energy difference between the ground
and first excited multiplets of the $i$-th bond. As $J_{i}>0$, it
can be shown that the ground state multiplet is represented by a vertical
Young tableau with $\widetilde{Q}$ lines, where $\widetilde{Q}=Q_{i}+Q_{i+1}$,
if $Q_{i}+Q_{i+1}\le N$, and $\widetilde{Q}=Q_{i}+Q_{i+1}-N$, if
$Q_{i}+Q_{i+1}>N$. The energies of ground and excited multiplets
can be calculated from the Casimir's with the usual trick $2\bm{\Gamma}_{i}\cdot\bm{\Gamma}_{i+1}=\left(\bm{\Gamma}_{i}+\bm{\Gamma}_{i+1}\right)^{2}-\bm{\Gamma}_{i}^{2}-\bm{\Gamma}_{i+1}^{2}$.
The value of the Casimir's of the relevant tableaux is given in Ref.~\cite{colemanpepintsvelik}.
We then decimate that bond by keeping only the ground multiplet and
renormalizing the neighboring interactions in the following fashion.
If $Q_{i}+Q_{i+1}=N$, the bond ground state is a singlet and is thus
removed from the system. The new effective coupling between the neighboring
spins $\bm{\Gamma}_{i-1}$ and $\bm{\Gamma}_{i+2}$ is, by second-order
perturbation theory,\begin{equation}
\widetilde{J}=\frac{2Q_{i}Q_{i+1}J_{i-1}J_{i+1}}{N^{2}\left(N-1\right)J_{i}}.\label{eq:renorm2}\end{equation}
 Otherwise, the spin pair $\bm{\Gamma}_{i}$ and $\bm{\Gamma}_{i+1}$ is replaced
by a new effective spin $\widetilde{\bm{\Gamma}}$, which belongs to a
totally antisymmetric representation with $\widetilde{Q}$ lines as
given above. It connects to the spins $\bm{\Gamma}_{i-1}$ and $\bm{\Gamma}_{i+2}$
through renormalized couplings given in \emph{first-order} perturbation
theory by\begin{equation}
\widetilde{J}_{i-1}=\xi_{i}J_{i-1},\textrm{ and }\widetilde{J}_{i+1}=\left(1-\xi_{i}\right)J_{i+1},\label{eq:renorm1}\end{equation}
 respectively, where $\xi_{i}=Q_{i}/\left(Q_{i}+Q_{i+1}\right)$,
if $Q_{i}+Q_{i+1}<N$, and $\xi_{i}=\left(N-Q_{i}\right)/\left(2N-Q_{i}-Q_{i+1}\right)$,
otherwise. We point out the similarity with the case of the random
chain with both ferro- and antiferromagnetic interactions \cite{westerbergetal},
where both 1st and 2nd order decimations are generated. Unlike the
latter, however, here the active (i.e., not yet decimated) spin clusters
are always vertical tableaux and the procedure always maintains a
totally antisymmetric spin chain. Moreover, the renormalized couplings
are always smaller than the original ones. Thus, at every decimation
step, the energy scale $\Omega$ is lowered.

An important feature of the decimation procedure is that it does not
privilege any specific representation. When a spin pair ($Q_{i},\, Q_{i+1}$)
is decimated out, the new effective spin is never equal to any of
the original ones ($\widetilde{Q}\neq Q_{i},\,\widetilde{Q}\neq Q_{i+1}$).
Thus, after an initial transient, each one of the $N-1$ totally antisymmetric
representations is equally probable, \emph{even if we start with}
$Q_{i}=Q=\mathrm{const.}$ (except for some special fine-tuned cases
dealt with later). We have confirmed this numerically, as will be
shown later. We have also checked that the distribution of representations
becomes uncorrelated with the distribution of couplings. We thus focus
on the flow of the coupling distribution, $P\left(J\right)\equiv P\left(J;\Omega\right)$,
as the highest scale $\Omega$ is decreased \cite{madasguptahu,fisherrandomchain}.
We take $\Omega=1$ initially.

As will be shown later, similarly to the random spin-$\frac{1}{2}$
chain, $P\left(J\right)$ always flows to an extremely broad distribution.
We are thus justified in neglecting the numerical prefactors in Eqs.~(\ref{eq:renorm2})
and (\ref{eq:renorm1}), which are always less than unity and irrelevant
asymptotically \cite{fisherrandomchain}. Furthermore, there are a
total of $\left(N-1\right)²$ possible decimation processes, all of
them equally likely. Of these, $N-1$ are 2nd order, each with probability
$p=1/\left(N-1\right)$, and the others are 1st order, with probability
$q=1-p$. Thus, we can write a flow equation in the useful logarithmic
variables $\Gamma=-\ln\Omega$, and $\zeta=\ln\left(\Omega/J\right)$
\cite{fisherrandomchain}

\begin{equation}
\frac{\partial}{\partial\Gamma}\rho(\zeta;\Gamma)=\frac{\partial}{\partial\zeta}\rho(\zeta;\Gamma)+q\rho(\zeta;\Gamma)\rho^{0}+p\rho^{0} \rho \otimes \rho,\label{eq:flow}
\end{equation}
where $\rho(\zeta;\Gamma)\textrm{d}\zeta=P\left(J;\Omega\right)\textrm{d}J$,
$\rho^{0}=\rho(0;\Gamma)$, and $\rho \otimes \rho =\int\textrm{d}\zeta_{1}\textrm{d}\zeta_{3}\rho(\zeta_{1};\Gamma)\rho(\zeta_{3};\Gamma)\delta\left(\zeta-\zeta_{1}-\zeta_{3}\right)$.
The first term on the right-hand side is due to the fact that $\zeta$
changes when $\Gamma$ increases. The second one, absent in the random
spin-$\frac{1}{2}$ chain, is due to 1st order decimation steps and
only ensures the normalization of $\rho$. The last one is due to
2nd order steps, which strongly renormalize $\rho$ broadening it.

If $P_{0}(J)$ is not extremely singular, the flow Eq.~(\ref{eq:flow})
has only one stable fixed point solution \cite{fisherrandomchain}
\begin{subequations}\label{eq:fixpt}\begin{eqnarray}
\rho^{*}(\zeta;\Gamma) & = & \frac{\theta(\zeta)}{p\Gamma}e^{-\zeta/(p\Gamma)},\\
P^{*}(J;\Omega) & = & \frac{\alpha}{\Omega}\left(\frac{\Omega}{J}\right)^{1-\alpha}\theta(\Omega-J),\end{eqnarray}
 \end{subequations}with $\alpha=1/(p\Gamma)=-\left(N-1\right)/\ln\Omega$.
The fixed point distribution (\ref{eq:fixpt}) broadens indefinitely
in the limit $\Omega\rightarrow0$, rendering the renormalization
procedure increasingly more precise, and asymptotically exact \cite{fisherrandomchain}.
The system is thus governed by an IRFP.

The relation between energy and length scales can be determined by
finding the fraction of active spin clusters $n_{\Gamma}$ at the
energy scale $\Gamma$ \cite{fisherrandomchain}. If $\Gamma$ is
increased by $\textrm{d}\Gamma$, a fraction
$dn_{\Gamma}=\left(2p+q\right)n_{\Gamma}\rho(0;\Gamma)\textrm{d}\Gamma $
of active spin clusters is decimated. Thus, close to the fixed point,
where $\rho(0;\Gamma)\approx\rho^{*}(0;\Gamma)$ \cite{fisherrandomchain}\begin{equation}
L_{\Gamma}\sim n_{\Gamma}^{-1}\sim\Gamma^{1/\psi}=\left[\ln\left(1/\Omega\right)\right]^{1/\psi},\label{eq:ngamma}\end{equation}
 where $\psi=p/(p+1)=1/N$. This type of `activated' dynamical
scaling, corresponding to a dynamical exponent $z\to\infty$, arises
here with an unexpected exponent $\psi$. When $N=2$, we recover the usual
form found in the random spin-$\frac{1}{2}$ chains \cite{fisherrandomchain,fishertransising}.

\begin{figure}
\begin{center}\includegraphics[%
  width=2.5in,
  keepaspectratio]{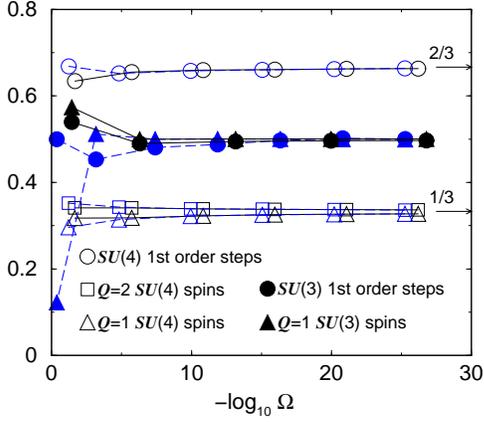}\end{center}

\caption{\label{cap:fig1}The fractions of first order decimation steps, of
spins in the fundamental ($Q=1$), and in the self-conjugate ($Q=2$,
only for $SU(4)$) representations, all as a function of $\Omega$.
For clarity, we only show data for chains A (solid lines) and E (dashed
lines) (see text). The filled (open) symbols refer to the $SU(3)$
($SU(4)$) chains. The data error is about the size of the symbol.}
\end{figure}

In order to check the validity of the approximations leading up to
Eq.~(\ref{eq:flow}), we have numerically implemented the full procedure.
The data were generated by decimating chains with lengths up to $10^{7}$,
averaging over 100 realizations of disorder. All the initial spins
belong to the fundamental representation ($Q_{i}=1$, $\forall i$).
We analyzed several initial distributions $P_{0}\left(J\right)$.
Chains A, B, and C had uniform distributions in the interval $x\leq J\leq1$,
with $x=0.9$, $0.5,$ and $0$, respectively. In chains D, E, and
F we used initial power-law distributions $P_{0}\sim J^{-\beta}$,
with $\beta=0.3$, $0.6,$ and $0.9$, respectively. In Fig.~\ref{cap:fig1},
we show the fraction of first order decimation steps as a function
of the energy scale $\Omega$, for the symmetry groups $SU(3)$ and
$SU(4)$. As anticipated, it tends asymptotically to $q=\left(N-2\right)/\left(N-1\right)$.
The figure also shows the fraction of $Q=1$ spins in the two cases,
and the fraction of self-conjugate ($Q=2$) spins in the $SU(4)$
chain. They all tend asymptotically to $1/\left(N-1\right)$, as expected.

\begin{figure}
\begin{center}\includegraphics[%
  width=2.5in,
  keepaspectratio]{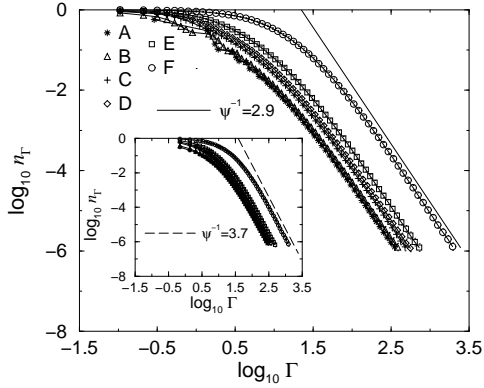}\end{center}

\caption{\label{cap:fig2}Fraction of active spins $n_{\Gamma}$ as a function
of the energy scale $\Gamma$, for the $SU(3)$ (main) and $SU(4)$
(inset) chains A to F. The data error is smaller than the symbol size. }
\end{figure}

In Fig.~\ref{cap:fig2}, we plot $n_{\Gamma}$ as a function of $\Gamma$
for $SU(3)$ and $SU(4)$. By fitting the asymptotic behavior, we
confirm the universality of the exponent $\psi=1/N$, as predicted
by Eq.~(\ref{eq:ngamma}). We point out that $\psi$ converges in
a logarithmic manner, thus a more precise determination of $\psi$
demands the decimation of longer chains than the ones studied here.
We see that as $N$ increases, so does the number of decimations needed
for a given decrease in energy scale. This `delayed scaling' can be
understood by realizing that only 2nd order processes are effective
in lowering the energy scale, and these become less frequent as $N$
increases.

There are other IRFP's in addition to the one analyzed above. For
example, the self-conjugate $SU(2k)$ spin chain (with integer $k>1$)
flows towards an IRFP with $\psi=1/2$, since $Q_{i}=k$, $\forall i$,
and only 2nd order decimation steps occur, like in the random spin-$\frac{1}{2}$
chain. Although these chains are gapful \cite{afflecksu2n}, they
are unstable against the introduction of weak disorder, due to the
topological nature of their ground state, as explained for the random
$J_{1}-J_{2}$ Heisenberg chain in Ref.~\cite{yangetal}. More importantly,
this $\psi=1/2$ IRFP is unstable against the introduction of $Q\neq k$
spins. For a small concentration $n_{i}$ of such spins, the system
will initially be governed by the $\psi=1/2$ IRFP, until the energy
scale $\Gamma\sim n_{i}^{-1/2}$ is reached. Below that scale, the
renormalization flow veers towards the IRFP of Eq.~(\ref{eq:fixpt}),
with the characteristic exponent $\psi=1/2k$. Similar IRFP's exist
for other $SU\left(N\right)$ chains, but they are equally unstable
with respect to the introduction of `defect' spins.

We can easily calculate the asymptotic behavior of thermodynamic quantities
using Eq.~(\ref{eq:ngamma}) \cite{fisherrandomchain}. Since $P(J)$
becomes very broad at low energies, the active spins are approximately
free at a low temperature $T=\Omega$, whereas the decimated ones
do not contribute, since they are frozen in singlet states with excitation
energies much greater than $T$. Hence, we find that the entropy density
$\sigma\sim n_{\Gamma}\sim\left(-\ln T\right)^{-1/\psi}$ and the
specific heat $c\sim\left(-\ln T\right)^{-(\psi+1)/\psi}$.
Furthermore, it can be easily shown that the magnetic susceptibility
of a single $SU\left(N\right)$ spin is Curie-like, from which it
follows that for the whole system $\chi\left(T\right)\sim n_{\Gamma}/T=1/[T\left(-\ln T\right)^{1/\psi}]$.

We can also obtain the asymptotic behavior of the spin correlation
function $C_{ij}\equiv\left\langle \bm{\Gamma}_{i}\cdot\bm{\Gamma}_{j}\right\rangle $
\cite{fisherrandomchain}. Spins belonging to the same cluster develop
$\mathcal{O}\left(1\right)$ correlations; otherwise, they are weakly
correlated. Therefore, such spins dominate the mean correlation function.
To find $\overline{C_{ij}}$, we need the probability that any two
well-separated spins $\bm{\Gamma}_{i}$ and $\bm{\Gamma}_{j}$ are rigidly locked
in the same spin cluster $\widetilde{\bm{\Gamma}}$ when $\left|i-j\right|\sim n_{\Gamma}^{-1}$.
First, we need to find $\mathcal{P}\left(t;n_{\Gamma}\right)$, the
probability to find a spin cluster $\widetilde{\bm{\Gamma}}$ composed
of $t$ original spins at scale $n_{\Gamma}$. After many decimations,
any spatial correlations between $Q$'s and $J$'s have vanished and
any remaining bond is equally likely to be decimated. The fraction
of clusters with $t$ spins at scale $n_{\Gamma}$ is $n_{\Gamma}\mathcal{P}\left(t;n_{\Gamma}\right)$.
When $dN_{dec}$ decimations are performed, $n_{\Gamma}$ decreases
by $dn_{\Gamma}=-\left(2p+q\right)dN_{dec}$ and\[
d\left[n_{\Gamma}\mathcal{P}\left(t\right)\right]=\left[-2\mathcal{P}\left(t\right)+q\sum_{t_{1},t_{2}}\mathcal{P}\left(t_{1}\right)\mathcal{P}\left(t_{2}\right)\delta_{t_{1}+t_{2},t}\right]dN_{dec},\]
where the two terms on the right-hand side give the fraction of decimated
and added clusters with $t$ spins and we suppressed the $n_{\Gamma}$
dependence of $\mathcal{P}$ to lighten the notation. Hence, \begin{equation}
n_{\Gamma}\frac{\partial\mathcal{P}\left(t\right)}{\partial n_{\Gamma}}=\frac{1-p}{1+p}\left[\mathcal{P}\left(t\right)-\sum_{t_{1}}\mathcal{P}\left(t_{1}\right)\mathcal{P}\left(t-t_{1}\right)\right],\label{eq:p_flow}\end{equation}
 whose solution is $\mathcal{P}\left(t;n_{\Gamma}\right)\sim n_{\Gamma}^{\gamma}\exp\left(-tn_{\Gamma}^{\gamma}\right)$
in the limit $n_{\Gamma}\rightarrow0$, with $\gamma=\left(1-p\right)/(1+p)=1-2/N$.
Finally, the probability that $\bm{\Gamma}_{i}$ and $\bm{\Gamma}_{j}$ are
active in the same cluster is equal to $\left(\overline{t}n_{\Gamma}\right)^{2}\sim\left(n_{\Gamma}^{1-\gamma}\right)^{2}$,
yielding,

\begin{equation}
\overline{C_{ij}}\sim\frac{\left(-1\right)^{i-j}}{\left|i-j\right|^{\phi}},\label{eq:mean_corr}\end{equation}
 with $\phi=4/N$. The typical correlation function, however, is very
different. Following \cite{fisherrandomchain}, we note that it involves
many factors of $\widetilde{J}$ decimated at various scales $e^{-\Gamma}$.
The scaling behavior is dominated by the smallest factor $\mathcal{O}\left(e^{-k\left|i-j\right|^{\psi}}\right)$, where $k \sim \mathcal{O}(1)$
yielding typical correlations\begin{equation}
\left|C_{ij}\right|_{typ}\sim\exp\left(-k\left|i-j\right|^{\psi}\right).\label{eq:typ_corr}\end{equation}

\begin{figure}
\begin{center}\includegraphics[%
  width=2.5in,
  keepaspectratio]{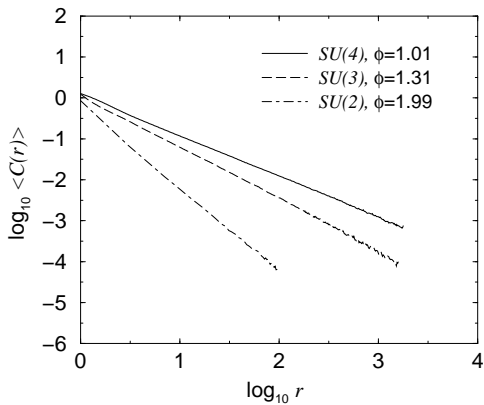}\end{center}

\caption{\label{cap:fig3}Mean correlation function for groups $SU(2)$, $SU(3)$
and $SU(4)$. The power-law dependence is evident. The exponents $\phi=1.99\pm0.03$,
$1.31\pm0.03$, and $1.01\pm0.03$ are obtained by fitting the region
$r>10$, and are in excellent agreement with the analytical value
$\phi=4/N$ (see Eq.~(\ref{eq:mean_corr})).}
\end{figure}

Fig.~\ref{cap:fig3} shows numerical results for the mean correlation
function $\overline{C_{ij}}$ for groups $SU(2)$, $SU(3)$ and $SU(4)$,
averaged over 200 realizations of disorder for chain lengths up to
$L=10^{5}$ and open boundary conditions. The numerical procedure
consists in completely decimating a chain, and counting the fraction
of spin pairs that become strongly correlated at the distance $\left|i-j\right|$
\cite{yusufyang1}. Excellent agreement with the analytical prediction
of $\phi$ is obtained. No significant dependence on initial disorder
strength was observed, confirming the universal behavior.

In the large-$N$ limit, the mean correlation function decays extremely
slowly. In this limit, the fraction of 2nd order processes is very
small and the mean number of spins in a cluster diverges at low energies,
all of them being strongly correlated. A $1/N$ expansion of Eq.~(\ref{eq:mean_corr})
leads to $\overline{\left|C_{ij}\right|}\sim1/\ln\left(\left|i-j\right|\right)$.
Incidentally, this is the same behavior observed numerically in random
ferro- and antiferromagnetic spin chains \cite{hikiharaetal}. This
is no surprise, since both systems are dominated by similar 1st order
decimations \emph{whose clustering rules are the same as} $N\to\infty$.
Therefore, they are both described by Eq.~(\ref{eq:p_flow}) with
$p=0$, hence the logarithmic dependence of the mean correlation function.
As far as we know, this analytical explanation has not appeared before.
However, we should stress that the asymptotic region governed by the
IRFP is reached at energy scales which decrease with the increase
of $N$, since the 2nd order processes become increasingly rare. Therefore,
in the infinite-$N$ limit the universal behavior described above
disappears and a direct infinite-$N$ approach fails to capture the
physics at any finite $N$.

Interestingly, some multicritical points of random antiferromagnetic
spin $S$ chains have been shown to exhibit a structure that is very
similar to the generic $SU(N)$ IRFP described above
\cite{damlehuse}. In particular, the energy-length scale exponent is
the same. In that case, $N$ is the number of phases meeting at the
multicritical point.

In conclusion, we have identified in random antiferromagnetic $SU\left(N\right)$
chains an infinite randomness fixed point with exponents different
than the ones previously found in spin-$\frac{1}{2}$ chains. An important
question which we leave for future study is the stability of this
phase against the introduction of anisotropy.

We acknowledge the financial support of FAPESP grants 01/00719-8 (E.M.),
03/00777-3 (J.A.H.) and CNPq grant 301222/97-5 (E.M.). We thank D. A. Huse for bringing Ref.~\cite{damlehuse} to our attention.

\end{document}